# Ultralow-pressure mechanical-motion switching of ferroelectric polarization

Baoyu Wang[1,#], Xin He[2,1,#], Jianjun Luo[3], Yitong Chen[4], Zhixiang Zhang[1], Ding Wang[1], Shangui Lan[1,2], Peijian Wang[1], Xun Han[1], Yuda Zhao[1], Zheng Li[1], Huan Hu[5], Yang Xu[1], Zhengdong Luo[6,7], Weijin Hu[8], Bowen Zhu[4], Jian Sun[9]*, Yan Liu[6,7], Genquan Han[6,7], Xixiang Zhang[10], Bin Yu[1]*, Kai Chang[2], and Fei Xue[2,1]*

[1] College of Integrated Circuits, ZJU-Hangzhou Global Scientific and Technological Innovation Center, Zhejiang University, Hangzhou 311215, China
[2] Center for Quantum Matter, School of Physics, Zhejiang University, Hangzhou 310027, China.
[3] CAS Center for Excellence in Nanoscience, Beijing Key Laboratory of Micro-nano Energy and Sensor, Beijing Institute of Nanoenergy and Nanosystems, Chinese Academy of Sciences, Beijing 100083, China
[4] Key Laboratory of 3D Micro/Nano Fabrication and Characterization of Zhejiang Province, School of Engineering, Westlake University, Hangzhou 310024, China
[5] ZJUI Institute, International Campus, Zhejiang University, Haining 314400, China
[6] Hangzhou Institute of Technology, Xidian University, Hangzhou 311200, China
[7] State Key Discipline Laboratory of Wide BandGap Semiconductor Technology, School of Microelectronics, Xidian University, Xi'an 710071, China
[8] Shenyang National Laboratory for Materials Science, Institute of Metal Research, Chinese Academy of Sciences (IMR, CAS), Shenyang 110016, China
[9] School of Physics, Central South University, Changsha 410083, China
[10] Physical Science and Engineering Division, King Abdullah University of Science and Technology, Thuwal 23955-6900, Saudi Arabia.
* Corresponding emails: jian.sun@csu.edu.cn; yu-bin@zju.edu.cn; xuef@zju.edu.cn
[#] These authors contributed equally.

**Ferroelectric polarization switching, achieved by mechanical forces, enables the storage of stress information in ferroelectrics, and holds promise for human-interfacing applications. The prevailing mechanical approach is locally induced flexoelectricity with large strain gradients. However, this approach usually requires huge mechanical pressures, which greatly impedes device applications. Here, we report an approach of using triboelectric effect to mechanically, reversibly switch ferroelectric polarization across α-$In_2Se_3$ ferroelectric memristors. Through contact electrification and electrostatic induction effects, triboelectric units are used to sensitively detect mechanical forces and generate electrical voltage pulses to trigger α-$In_2Se_3$ resistance switching. We realize multilevel resistance states under different mechanical forces, by which a neuromorphic stress system is demonstrated. Strikingly, we achieve the reversal of α-$In_2Se_3$ ferroelectric polarization with a record-low mechanical pressure of ~ 10 kPa, and even with tactile touches. Our work provides a fundamental but pragmatic strategy for creating mechanical-tactile ferroelectric memory devices.**

**Teaser sentence:** Ultralow-pressure mechanical forces reversibly switch ferroelectric polarization, enabling neuromorphic tactile devices.

# INTRODUCTION

Mechanical switching of ferroelectric polarization has gained considerable attention because mechanical forces provide a new knob beyond electric fields to control ferroelectric properties (*1-6*). Generally, using piezoresponse force microscopy (PFM) probes, mechanical forces can be locally applied onto nanoscale circular regions of ferroelectric crystals, i.e., ~20 nm in radius. The non-uniform force leads to a large strain gradient along the direction of mechanical force and commonly induces flexoelectric fields across the crystal (Fig. 1A top), which can consequently tilt the double-well energy landscape and align ferroelectric polarization to a stable state (*1, 7, 8*). To date, flexoelectricity-mediated ferroelectric polarization reversal has been revealed in many oxide ferroelectrics (*9-13*), and the sample thicknesses range from micrometer to nanometer scale (*5, 13-17*). These works demonstrate the universality and effectiveness of the mechanical approach, which shows great potential for practical use, e.g., tactile ferroelectric memristors.

Despite the prospective applications, creating tactile memristors based on mechanical control over ferroelectric polarization still presents challenges. This is because, for switching polarization and storing stress information, the device usually requires a huge mechanical pressure of around ~ GPa at nanoscale areas such that a required strain gradient-induced internal electric fields (i.e., flexoelectric fields) can emerge within the channel material. To overcome this difficulty and promote device development, a fundamentally different approach, capable of reversing polarization at a tactile-pressure range (~ kPa), is highly desired. Here, we use triboelectricity, instead of flexoelectricity, to generate sizable electric fields and reversibly switch ferroelectric polarization (Fig. 1A). In response to mechanical stimulus, triboelectricity effect under a pressure of as low as ~kPa can produce a broad range of electric fields via the delicate selection of triboelectric materials and interface design. As shown in Fig. 1B, although both mechanical methods involve the transformation of applied stimulus into electric fields for toggling ferroelectric polarization, our approach sidesteps several practical limitations of flexoelectricity counterpart, such as the requirement of ultrahigh pressure, local polarization reversal, and unidirectional switching. By adopting triboelectricity we achieve a record-low mechanical pressure (i.e., 1~10 kPa) to reverse the polarization and incrementally tune resistance switching. Multilevel resistance states towards neuromorphic stress systems are demonstrated. Our work provides a practical way to build ferroelectric stress memristors applicable to human-interfacing applications.

To better understand the physics behind triboelectricity-induced polarization switching, we plot a potential energy well related to triboelectricity, and a Landau double-well energy landscape associated with ferroelectricity on the same vertical axis (*E*), as shown in Fig. 1C. In the initial state (top panel of Fig. 1C), materials A and B with different energy wells are physically separated, and ferroelectric polarization is supposed to orientate upwards. When these two triboelectric materials are in mechanical contact electrification (bottom panel of Fig. 1C), electron transfer occurs at the contact interface due to the strongly overlapped electron clouds, resulting in the emission of energy $E_T$ that is equal to the energy difference $E_A$ - $E_B$ (*18, 19*). By choosing desirable triboelectric materials and engineering their interfaces, the $E_T$ energy can be tuned such that enough energy can be converted to drive polarization switching. By contrast, in the classic Landau theory, the free energy of ferroelectric materials is generally described by a double energy well landscape (middle panel of Fig. 1C), wherein two degenerate energy minima represent two stable spontaneous polarization states, -P and +P. An energy barrier (i.e., $E_F$) exists between these two states but can be overcome by the application of external stimuli with energy higher than $E_F$ (*20-22*). Therefore, when $E_T$ > $E_F$ upon mechanical stimulus, ferroelectric polarization switching can be initiated. Under the application of an $E_T$-induced negative bias, the Landau double well is destabilized, promoting downward polarization switching. Conversely, when an $E_T$-induced positive bias is applied, the polarization switches upwards. Note that $E_T$ can be converted into over kilovolt-voltage and milliampere-current outputs through triboelectricity-based energy devices (i.e., triboelectric units) (*23, 24*), delivering a spectrum of electric field magnitudes (the red rectangle in Fig. 1D). Therefore, the $E_T$ can easily exceed the energy of the coercive fields of commonly studied ferroelectric materials (*5, 25-30*). The detailed analysis of underlying physical mechanisms can be found in Supplementary Text S1.

## RESULTS

### Mechanically switching ferroelectric polarization

To demonstrate mechanical triboelectricity control over ferroelectric polarization, we adopt free-standing (sliding) triboelectric devices as perceiving units for converting mechanical stimuli into electric voltages (Fig. 2A). In the triboelectric unit (the bottom of Fig. 2A), polytetrafluoroethylene (PTFE) polymer with electronegativity and copper film with electropositivity were chosen as triboelectric materials to accommodate the energy requirement of global polarization switching. The bottom component of triboelectric unit is a fixed rectangular block made up of three-material stacking: acrylic, copper foil, and PTFE, while the top component is a movable square comprising acrylic, sponge and copper foil. Apart from these triboelectric units, we constructed planar ferroelectric memristor and monitor mechanical polarization switching and resulting resistance variation on a device level. In ferroelectric devices, exfoliated ferroelectric α-$In_2Se_3$ flakes serve as active semiconductor channels with lengths of around 1.5 μm, and Ti/Au bilayer electrodes were used to form Schottky contacts for realizing high on-off ratios (*31-34*). Raman spectra acquired from α-$In_2Se_3$ flakes (fig. S2) exhibits strong peaks at 90, 103, 180, and 195 cm$^{-1}$, which are respectively attributed to the E, A(LO+TO), A(TO), and A(TO) phonon modes (*35, 36*). Notably, these peaks coincide well with hexagonal (2H) α-$In_2Se_3$ phase (*31, 37*). In addition, to validate the α-$In_2Se_3$ ferroelectricity, we conducted local PFM hysteresis loops and acquired spontaneous domains (fig. S3 and S4). The results remarkably indicate the intrinsic switching of ferroelectric polarization under an external electric field. We have noticed the debate over α-$In_2Se_3$ in-plane ferroelectricity, which is discussed in Supplementary Text S2. Moreover, we note that, during device fabrication process, aqueous solutions are avoided to reduce the screening effect on polarization charges and improve memristor performance (*38*).

The operating principle of triboelectric units is based on the coupling effect of contact electrification and electrostatic induction between two solids (*39-42*), as illustrated in Supplementary Text S3. When the movable square slides across the bottom fixed component, charge transfer occurs and the open-circuit output voltages can be generated. The magnitude of output voltages is not only strongly dependent on the applied frictional force but also on the external load resistance in the circuit. Through matching the resistance difference between an oscilloscope and α-$In_2Se_3$ device, the output voltages from each friction increase with mechanical stress and can be effectively controlled within an applicable range of 0~80 V, as shown in Fig. 2B, which is obviously enough to modulate most ferroelectric polarization including α-$In_2Se_3$. With the appropriate load resistance, the specific output voltages for each friction are dependent on the magnitude of the applied mechanical force, allowing us to obtain the exact voltages required for controlling over polarization across α-$In_2Se_3$ ferroelectric memristors. The polarity of the output voltages can be simply determined by changing the connecting conditions of the triboelectric unit, i.e., forward and backward connections. From the enlarged inset of Fig. 2B, we observe that the output voltage exhibits a single pulse waveform with the pulse width of approximately 0.1 ms.

According to our previous work (*38*), resistance switching in α-$In_2Se_3$ ferroelectric memristors is closely associated with polarization switching across the channel, and thus can be employed to characterize mechanical modulation of ferroelectric polarization. Therefore, we first apply an electric field sweeping onto the memristor to assess the performance of resistance switching. Fig. 2C shows the electrical hysteresis loops of our α-$In_2Se_3$ ferroelectric memristor under diverse bidirectional sweeping voltages. It is seen that the on/off switching ratio increases with increasing maximum sweeping voltages and ultimately reaches ~$10^4$ at 26 kV/cm, in which the ferroelectric polarization has been completely switched. We note that the high on/off ratio could enable a multitude of intermediate resistance states through gradually modulating ferroelectric polarization, which is crucial for neuromorphic computing devices (*37, 43, 44*).

Mechanical-force-induced polarization switching is revealed by a series of resistance states in Fig. 2D. We choose small sweep electric fields (from -1.6 to 1.6 kV/cm) to read resistance states stimulated by mechanical forces. Upon mechanically sliding the triboelectric unit under forward connection, the resistance changes from the initial state (black curve) to the high resistance (blue curve), and then reversely switches to the low resistance (red curve) under backward connection. Resistance variation as a dependence of mechanically sliding motions was systematically explored and the results are shown in fig. S7, which demonstrates the characteristic of adjustable multilevel current states. These observations preliminarily suggest that triboelectricity mechanical sliding motions can realize ferroelectric polarization switching.

The exact mechanism behind mechanically switched resistance is involved with two physical processes. The first is that, via contact electrification and electrostatic induction, sliding triboelectric unit transform mechanical motions to electrical pulses, and the second process is that these generated electrical pulses modulate ferroelectric polarization and result in resistance variation, as illustrated in Supplementary Text S1. A more detailed analysis of the underlying mechanism of α-In$_2$Se$_3$ memristor resistance variation is shown in Supplementary Text S6. For comparing mechanical and electrical switching of ferroelectric polarization, we acquired resistance switching under the application of electrical voltage poling (Fig. 2E). Blue and red curves represent the currents after applying electric fields of +100 and -133 kV/cm, respectively, corresponding to the high and low resistance states. The current variation highly resembles that excited by mechanical force switching.

To visualize mechanical-sliding-motion-dependent polarization switching, we used the Dual AC Resonance Tracking (DART) mode of Oxford MFP-3D AFM to in-situ scan α-In$_2$Se$_3$ channel, wherein the orientation of ferroelectric polarization determines resistance states. As shown in Fig. 2F, mechanical motions under forward and backward connections can lead to amplitude and phase changes over the α-In$_2$Se$_3$ channel. During the measurement, the mechanically-modulated high resistance state, read by a small bias, exhibits a bright phase response while the low resistance state shows a dark response. We note that the phase contrast under different connections shall stem from different polarization orientations triggered by mechanical motions. Consistent with previous studies(*45-47*), the amplitude mapping also exhibits relevant changes (left panel of Fig. 2F) upon applying mechanical stimuli. For forward connection, α-In$_2$Se$_3$ polarization is compelled to point upwards as a response to mechanical motion. The positive and negative polarization charges can distribute at source and drain interfaces, tuning the interfacial energy band and giving rise to the high resistance state. By contrast, after being stimulated by triboelectric unit under backward connection, the ferroelectric polarization is fully reversed to P$_{down}$. Consequently, the conductance changes to a low resistance state due to the reduction of the Schottky barrier induced by downward energy band bending. To further demonstrate the polarization reversal, we derived exact phase response from the given blue and red lines in forward and backward phase mapping, and plotted them in fig. S8 for quantitative comparison. A striking contrast of above 100° can be seen, proving ferroelectric polarization reversal. The corresponding topography after poling is shown in fig. S9, which remains unchanged, thereby ruling out the possibility that the ferroelectric polarization switching stems from morphological changes.

**Neuromorphic stress device**

Having fundamentally established the concept that triboelectricity can modulate ferroelectric polarization, we attempt to develop neuromorphic stress memristors. Neuromorphic devices that are capable of simulating the neurons and synapses in our brain feature parallel sensing, processing, and storing of input information with high speed and low power consumption (*48*). Such a type of device is important to address high throughput data within a short timeframe and create artificial neural networks for artificial intelligence(*49, 50*). Here, we aim to integrate mechanical stress sensing and storage functionalities into α-In$_2$Se$_3$ ferroelectric memristors in a practical way to emulate the functions of biological mechanoreceptors and neuronal processing. Towards this aim, the pressing mode triboelectric unit is adopted to sense mechanical stress signals instead of the sliding mode counterpart because the pressing mode architecture favors stress sensing and its stress loads can be quantitatively measured by a force gauge. In our two-terminal α-In$_2$Se$_3$ ferroelectric memristors, input electrode serves as the presynaptic terminal under the stimulation of press force. This force induces a channel current variation, which is analogous to the postsynaptic current. The tunable multilevel conductance states correspond to neural synaptic plasticity, where the resistance values represent synaptic weights.

As shown in Fig. 3A, the stress-responsive pressing-mode triboelectric unit can biomimetically act as a mechanoreceptor (the operating principle is provided in Supplementary Text S3). Each external press force stimulus applied on the triboelectric unit can be converted into a pulse voltage, and can be also transmitted to α-In$_2$Se$_3$ memristor through bonding wires, which like artificial afferent nerve fibers. The postsynaptic current flowing through α-In$_2$Se$_3$ channel is sensitive to the stimulation of external press forces and the mechanical stimulus can be stored, akin to the neural synaptic plasticity and memory functions of cerebral cortex. It is worth noting that the triboelectric-unit-based mechanoreceptor can directly convert mechanical pressures to pulse voltages without a ring oscillator, making our stress system more concise compared to the previous report (*51*). Our measurement setup is described in Supplementary Text S4.

As seen in fig. S11, the output voltages of the pressing mode triboelectric unit increase with increasing press force and there is a quasi-linear relationship between them. Therefore, it is feasible to control the amplitude of the output voltage through applied press force. Fig. 3B depicts the output voltages applied in the tests as a function of the press force under forward and backward connections. Note that, the applied press force should be elaborately controlled, which ensures that the α-In$_2$Se$_3$ channel would not be burned down with much higher output triboelectric voltages. Fig. 3C and 3D show resistance switching tuned by different press force under forward and backward connections, respectively. In the forward scenario, the current decreases with the press forces, whereas in the backward scenario, the current increases with each press force stimulation. These intermediate resistance states can suggest stress-tunable polarization dynamics: ferroelectric domains across the channel are gradually switched from upward to downward direction.

A large press force can lead to a large resistance variation (fig. S12), analogous to the electrical modulation of resistance switching. This mechanical-to-electrical effect is also similar to biological mechanoreceptors that can respond to pressure and generate nerve impulses. In addition, to demonstrate the reproducibility of the current state variation, we modulated the resistance state of another device with continuous switching of forward and backward connections, as shown in fig. S13. The trend of current variation corresponds well with the connection direction, confirming the repeated controllability of the ferroelectric polarization at the channel. Because of the ability to regulate resistance switching, Fig. 3E depicts the retention properties of multilevel resistance states related to different press forces. It is seen that each resistance state can persist for a relatively long time (i.e., 250 seconds) owing to the relatively robust ferroelectric domain stability, which suggests that the applied press forces can be perceived and recorded in the device, resembling the memory function of the brain. The long-term retention of around 1500 s is also measured as shown in fig. S14.

A key characteristic of artificial synapses for neuromorphic computing is the pulse number-dependent long-term potentiation and depression (LTP and LTD), which can also be simulated by our device via both electrical and force-tunable polarization dynamics. The 50 consecutive pulse voltages with amplitude of ±2 V and width of 10 ms were applied to α-In$_2$Se$_3$ device, and the read voltage of 0.5 V was set to read currents after the stimulation of each pulse voltage. As shown in Fig. 3F, the postsynaptic currents flowing through α-In$_2$Se$_3$ channel show remarkable potentiation and depression behaviors. We also applied similar mechanical press of approximately 0.5 N to progressively excite the synaptic memristor under backward connection while inhibiting it under forward connection (Fig. 3G). The method for measuring LTP/LTD characteristics under mechanical press stimulations is different from that under electrical pulse voltages. The detailed measurement process is provided in Supplementary Text S5. Moreover, to highlight the tunability and flexibility of synaptic weight modulation, we applied electrical and mechanical stimuli for depression and potentiation characteristics, respectively. As depicted in Fig. 3H, the postsynaptic current is initially inhibited by positive electrical pulse voltage and subsequently excited by mechanical stress, demonstrating an additional degree of modulation freedom.

In addition, symmetrical LTP and LTD as well as multilevel states are essential for accelerating neural network training (*52, 53*). We constructed an artificial neural network (ANN) that employs the LTP and LTD characteristics under electrical and mechanical stimulations for recognizing 28 × 28-pixel handwritten digits image in the MNIST dataset (Fig. 3I). We used equations 1 and 2 (where *ΔG* is the variation of synaptic weight, and $G_{n+1}$ and $G_n$ represent synaptic weight after $n+1_{th}$ and $n_{th}$ modulation) respectively to fit the LTP/LTD characteristics (*54*), obtaining parameters of $α_P$, $β_P$, $α_D$, and $β_D$. Based on these fitted parameters, we then defined a strict update rule for synaptic weights that is governed by equations 1 and 2.

$$G_{n+1} = G_n + \Delta G_P = G_n + \alpha_P e^{\beta_P (G_n - G_{min}/G_{max} - G_{min})} \qquad (1)$$

$$G_{n+1} = G_n + \Delta G_D = G_n - \alpha_D e^{\beta_D (G_{max} - G_n/G_{max} - G_{min})} \qquad (2)$$

As shown in Fig. 3J, after 100 training cycles, the accuracy rates fed with experimental LTP/LTD behaviors can reach 94.7% and 93.6%, respectively. These performances imply that our approach of triboelectricity-induced polarization holds immense potential for the development of neuromorphic stress systems towards bio-robot and electronic skin. A conceptual neuromorphic tactile system based on an α-In$_2$Se$_3$ memristor array is illustrated in fig. S16.

**Ferroelectric resistance switching realized by finger touch**

We believe that record-low mechanical pressures to switch polarization can be further optimized because the elasticity of pressing-mode triboelectric unit is largely restricted by polymer intrinsic elastic properties. If delicately designing the triboelectric unit, we anticipate that tactile touch can also induce polarization switching. In doing so, we use a single-electrode-mode triboelectric unit that is extremely sensitive to perceive tactile touch instead of pressing-mode counterparts. Fig. 4A illustrates the schematic of device architecture, the real photo is shown in fig. S17. Copper foil and PTFE film were sequentially adhered to the back side of Si substrate of α-$In_2Se_3$ devices. A 20-ohm resistor was selected as load resistance, which connects triboelectric unit with the channel via wire bonding. Our fingers, acting as free-moving objects, can directly touch copper foil and PTFE stacking to generate output voltages through contact electrification (see Supplementary Text S3 for triboelectric unit mechanism). The polarity of voltage can be switched through electrical connections.

Fig. 4B shows the corresponding pulse voltages when the finger touches the triboelectric unit under both forward and backward connections. Under forward connection, the device is stimulated by a positive pulse voltage upon the first finger tactile touch ($Touch_{1st}$), resulting in a decreased current (Fig. 4C). However, after the second and third tactile touch ($Touch_{2nd}$ and $Touch_{3rd}$) under backward connection, the device current increases owing to the stimulation by negative voltages (Fig. 4D). Moreover, the fourth and fifth tactile touches ($Touch_{4th}$ and $Touch_{5th}$) were carried out with forward and backward connections, respectively, and the resulting current first decreases and then increases (Fig. 4E and 4F). These results demonstrate that tactile touches can also reversibly switch ferroelectric polarization, which is distinguishable from flexoelectricity-induced single-direction switching (*4*). For comparison, the current responses as a function of different touches are plotted in fig. S18. It is worth noting that the small current variation is due to the relatively small output voltages generated by finger tactile touches.

We emphasize that for flipping α-$In_2Se_3$ ferroelectric polarization the applied stress pressure is estimated to be as low as 1~10 kPa, given that the magnitude and area of finger press force remain around 0.5~5 N and 0.5 × 0.5 $cm^2$, respectively. The low pressure of our triboelectricity approach (~ 10 kPa) is five orders of magnitude smaller than that of flexoelectricity-induced polarization switching (~GPa), as shown in Fig. 4G (*2, 5, 9, 55-58*).

This giant mechanical modulation can be attributed to the high voltage-output efficiency of our triboelectric units via electrification and electrostatic induction. Our results bring an applicable but high-efficiency way to apply the basic effect, i.e., mechanical switching of polarization switching, to the device level, which can be extended to other ferroelectric materials such as $BiFeO_3$ (fig. S23).

**DISCUSSION**

We have realized ultra-low pressures, even tactile touches, to mechanically switch ferroelectric polarization through triboelectricity, ingeniously bypassing the device limitation of the flexoelectric effect. The actual required stress pressure for our method is as low as 10 kPa, ensuring the possibility of being utilized in practical ferroelectric stress devices. By integrating a pressing-mode triboelectric unit with an α-$In_2Se_3$ ferroelectric memristor, we have created an artificial neuromorphic stress system, simultaneously realizing stress perception and storage functionalities with multi-level resistance states. Utilizing experimental LTP and LTD characteristics, we have also constructed an artificial neural network, demonstrating a high accuracy of image recognition.

## MATERIALS AND METHODS

### Device fabrication

2D α-$In_2Se_3$ flakes with the thickness of around 50 nm were obtained by mechanical exfoliation from bulk crystals. They were then transferred onto the heavily doped silicon substrates with a 285 nm $SiO_2$ layer through a 2D materials transfer stage. The heavily doped silicon also acts as the back gate in α-$In_2Se_3$ field effect transistors. Afterward, the two electrodes' pattern on the α-$In_2Se_3$ layer were defined by laser direct write lithography and electron beam lithography. And then Ti/Au (10 nm/40 nm) or Cr/Au (10 nm/40 nm) metals were deposited by thermal evaporation. The positive 5350 photoresist was lifted off through sequentially soaking in acetone and isopropanol solutions. It is worth mentioning that aqueous solutions were avoided during the whole microfabrication processes, as mentioned above, to reduce the screening effect on ferroelectric polarization charges. Subsequently, the triboelectricity unit was integrated with the prepared devices through wire bonding.

### Fabrication of triboelectric units

In our experiment, three distinct types of triboelectric units were utilized: free-standing (sliding) mode, pressing mode, and single-electrode mode. The sliding-mode triboelectric unit consists of bottom and top components, in which PTFE film and copper foil are actual triboelectric materials for generating electricity. The bottom component was made up of three materials: An acrylic board with a size of 17.5 × 5 $cm^2$ acts as a base; Two separated 150-μm-thick copper foils were adhered to the fixed base, serving as bottom electrodes; Subsequently, an 80-μm-thick PTFE film was adhered to copper foils. As for the top movable component, a 5-mm-thick sponge layer was attached to the acrylic board with a size of 4 × 5 $cm^2$ for better surface contact, and a 150-μm-thick copper foil was directly adhered onto the sponge layer.

The pressing-mode triboelectric unit is in an arched structure by using polydimethylsiloxane (PDMS) and indium tin oxide (ITO) as two triboelectric materials. A PDMS layer (Sylgard 184, Tow Corning) was formed by spin-coating and heating at 80 °C for 60 min, and subsequently adhered to a transparent indium tin oxide (ITO)-coated polyethylene terephthalate (PET) film. Such a PDMS-ITO-PET stacking acts as top component. Another ITO-coated PET film was placed onto an acrylic board, serving as bottom component. The top and bottom components were sealed at the two ends to form the arched structure.

As for single-electrode-mode triboelectric unit, a 150-μm-thick copper foil and 80 μm thick PTFE film were sequentially adhered to the back side of Si substrate in α-In$_2$Se$_3$ devices. Our fingers, acting as free-moving objects, can directly touch the copper foil and PTFE stack to generate triboelectricity.

**Electrical performance and PFM measurements**

The output voltages from the three modes triboelectric units were all measured through a digital oscilloscope (RIGOL MSO8104) with the internal load resistance set at 100 megaohms. The applied stresses were measured by a commercial gauge with a resolution of 0.1 N. Keithley 2636B source meter was used to measure the electrical properties of α-In$_2$Se$_3$ devices on a homemade probe station at a room-temperature ambient environment. The electrical pulse voltages were generated from Keithley 2636B. Atomic force microscope (AFM, Oxford MFP-3D) in tapping mode was used to characterize the device morphology. In-situ PFM measurement was performed in Dual Ac Resonance Tracking (DART) mode using an SCM-PIT-V2 type conductive tip with a spring constant of 3 N/m at a resonance frequency of approximately 280 kHz to acquire amplified piezoelectric responses.

**BiFeO$_3$ film growth and fabrication of BiFeO$_3$-baesd memristors**

High-quality epitaxial BiFeO$_3$ film with a thickness of 200 nm was deposited on SrTiO$_3$ (STO) single-crystal substrate with a 20-nm-thick SrRuO$_3$ (SRO) bottom layer by pulsed laser deposition with a KrF exciton laser (λ = 248 nm). Prior to film growth, the STO substrate underwent a two-step surface treatment: first, it was etched with a buffered oxide etching solution and annealed at 1000 °C for 2 hours to obtain a TiO$_2$-terminated surface. Subsequently, SRO films were deposited at 750 °C under an oxygen pressure of 10 Pa and a laser energy density of 1.5 J/cm² with a repetition rate of 3 Hz and were then annealed under the as-grown condition for 10 min. Finally, BiFeO$_3$ film was deposited at 700 °C under the same oxygen pressure and laser energy. To enhance the crystallinity of the BiFeO$_3$ film, an annealing treatment was performed under a large oxygen pressure of $4 \times 10^4$ Pa for 30 minutes. The fabrication process of BiFeO$_3$ memristors involves creating a 50×50 μm² region on the BiFeO$_3$ film by photolithography, followed by depositing a 50 nm thick Pt layer as top electrodes via magnetron sputtering. The SRO layer under BiFeO$_3$ film acts as a bottom electrode. Triboelectric unit was connected with the BiFeO$_3$-based memristor via wire bonding.

**Acknowledgements**

The authors appreciate the support of the ZJU Micro-Nano Fabrication Center and ZJU-Hangzhou International Innovation Center.

**Funding:** This work was supported by the National Natural Science Foundation of China (No. 62304202 to F. X., 62274128 and 92264202 to Z.-D. L., 62090033 to Y. L., 62025402 to G. H.), the Zhejiang Provincial Natural Science Foundation of China (grant no. LDT23F04013F04 to F. X. and LDT23F04023F04 to Z.-D. L.), and the National Key R&D Program of China (2022YFA1204900 to J. S.).

**Author contributions:** F. X. conceived and guided the research. B. W., Y. C., B. Z., S. L., and P. W. fabricated devices. J. L. provided triboelectric units. B. W. and D. W. tested the electrical performance of devices. B. W., X. H., and H. H. carried out the PFM measurements. F. X., B. W., X. He, and J. L. analyzed the data. Z.-X. Z. helped draw the schematic figures. F. X. and B. W. wrote the paper. X. Han, Y. Z., Z. L., and X.Y. provided experimental resources. J. S., Z.-D. L, W. H., Y. L., G. H., and K. C. helped improve the paper. F. X., B. Y., and X.-X. Z. supervised the project.

**Competing interests:** F. X., B. W., and B. Y. are inventors listed on the Chinese patent ZL202311208196.9 (granted date: 19 December 2023) held by ZJU-Hangzhou Global Scientific and Technological Innovation Center that covers a ferroelectric semiconductor device, a touch-sensing storage device, and the method for writing and reading tactile data. All other authors declare that they have no competing interests.

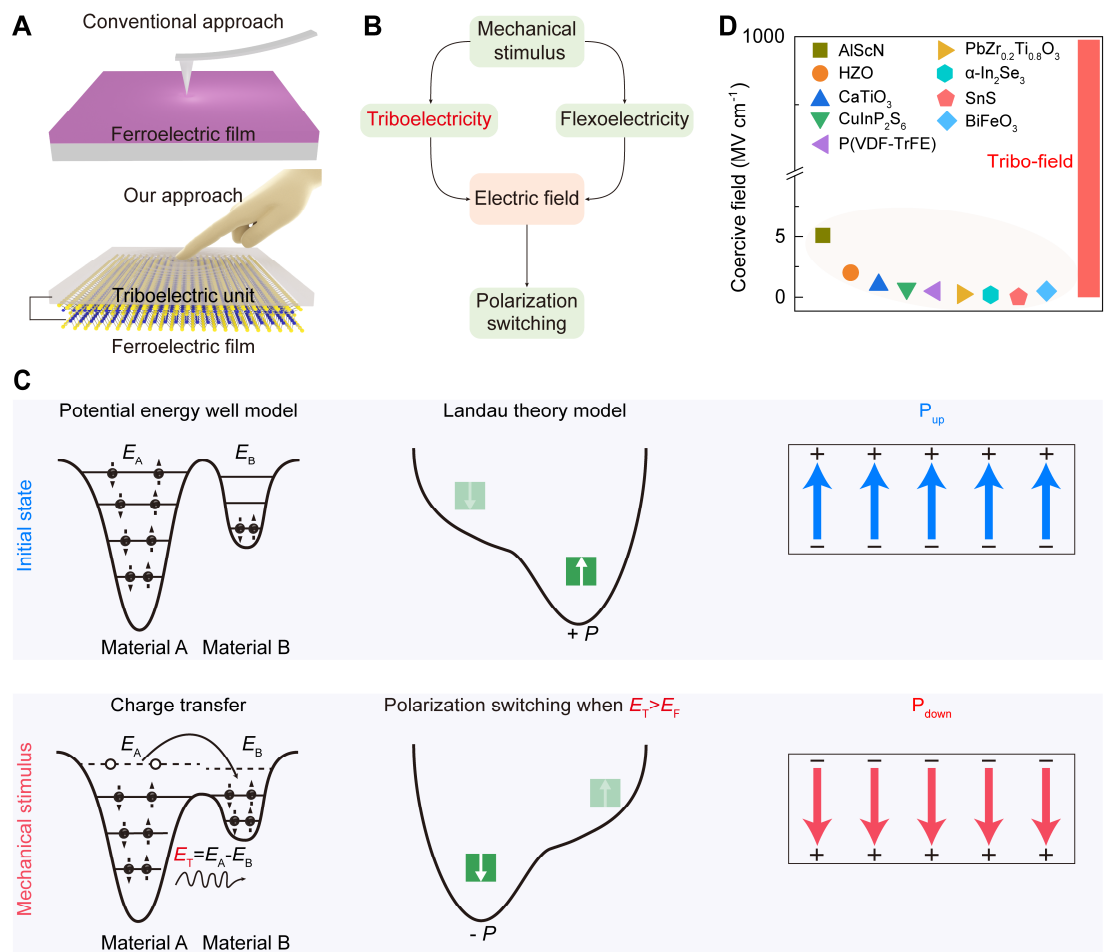

**Fig. 1. A novel approach to mechanically switch ferroelectric polarization at a global scale. (A)** Top panel: Illustration of the conventional, local mechanical-switching scheme, i.e., flexoelectricity-induced polarization reversal enforced by PFM probes. Bottom panel: By contrast, we propose an approach of triboelectricity to realize global mechanical switching of ferroelectric polarization. Note that triboelectric unit can be integrated on the ferroelectric film, allowing the direct application of mechanical stress. **(B)** Comparison of two mechanical approaches. Both of them require the transformation of electric field. **(C)** Potential energy well model for triboelectric effect and Landau double-well energy landscape for ferroelectric polarization. When two materials undergo contact electrification upon applying mechanical motion, charge transfer occurs, emitting energy $E_T$. With the energy $E_T$ much higher than the Landau energy barrier $E_F$, ferroelectric polarization switches from $P_{up}$ to $P_{down}$. Note that by changing the connection of the triboelectric unit, bidirectional switching can be achieved. **(D)** The electric field generated by the triboelectric effect (Tribo-field) can be much higher than the coercive fields of commonly studied ferroelectric materials.

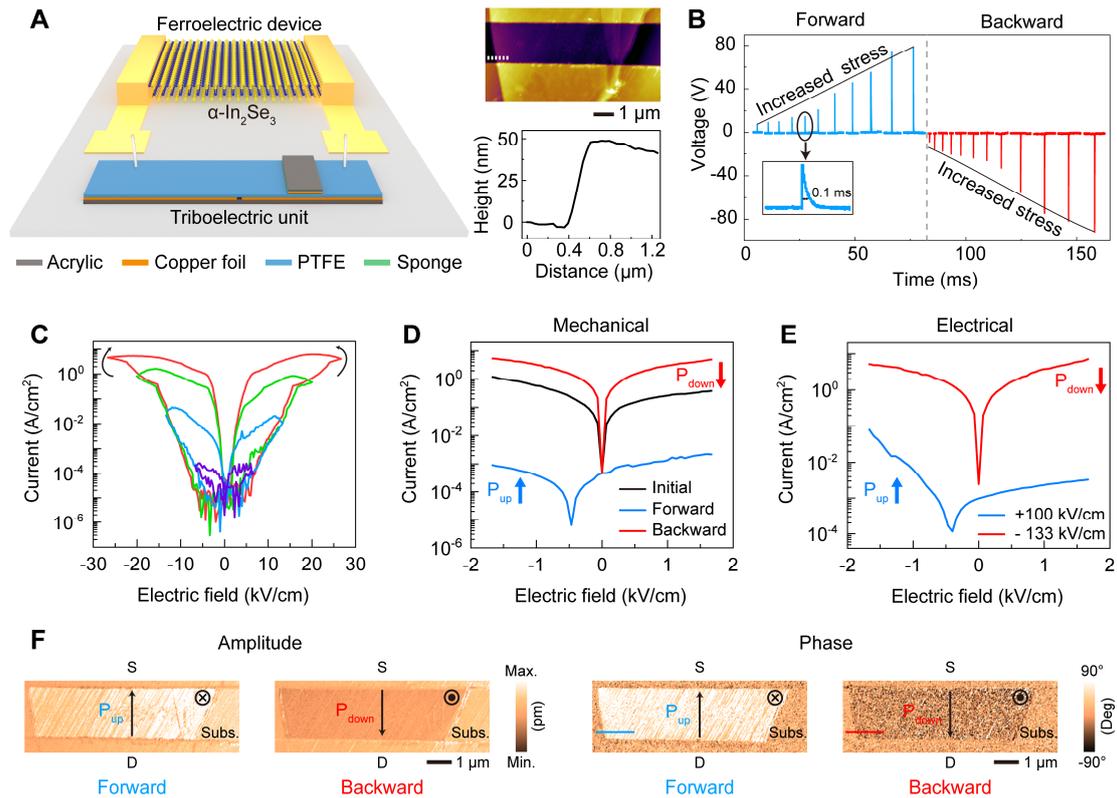

**Fig. 2. Characterization of triboelectricity-induced mechanical switching of ferroelectric polarization via electrical hysteresis and PFM amplitude mapping. (A)** Schematic of our experimental setup composed of α-In$_2$Se$_3$ ferroelectric memristor connected with a sliding mode triboelectric unit. All components of the triboelectric unit are clearly labeled with small squares. Atomic force microscope (AFM) topographic image and corresponding α-In$_2$Se$_3$ channel thickness of approximately 50 nm (right). **(B)** Typical output voltages for a sliding mode triboelectric unit with forward (blue curves) and backward (red curves) connection. The amplitude of the output voltage is increased with increased stress. **(C)** I-V electrical hysteresis loops under electric-field sweeping of 6, 14, 20, and 26 kV/cm, corresponding to purple, blue, green, and red curves. The arrows indicate the exact sweeping direction. **(D)** Resistance switching modulated by the triboelectric effect. The black curve is the initial current state, whereas the red and blue curves indicate the current states after applying forward and backward triboelectric voltages, respectively. **(E)** Electrically modulated resistance switching for comparison. The blue and red curves represent the currents after applying electric fields of +100 and -133 kV/cm, respectively. **(F)** Ferroelectric polarization switching probed by piezoelectric force microscope (PFM) amplitude and phase mapping under forward and backward connections. The arrows represent the in-plane polarization orientations while "⊕" and "⊙" indicate the out-of-plane polarization pointing inward and outward, respectively.

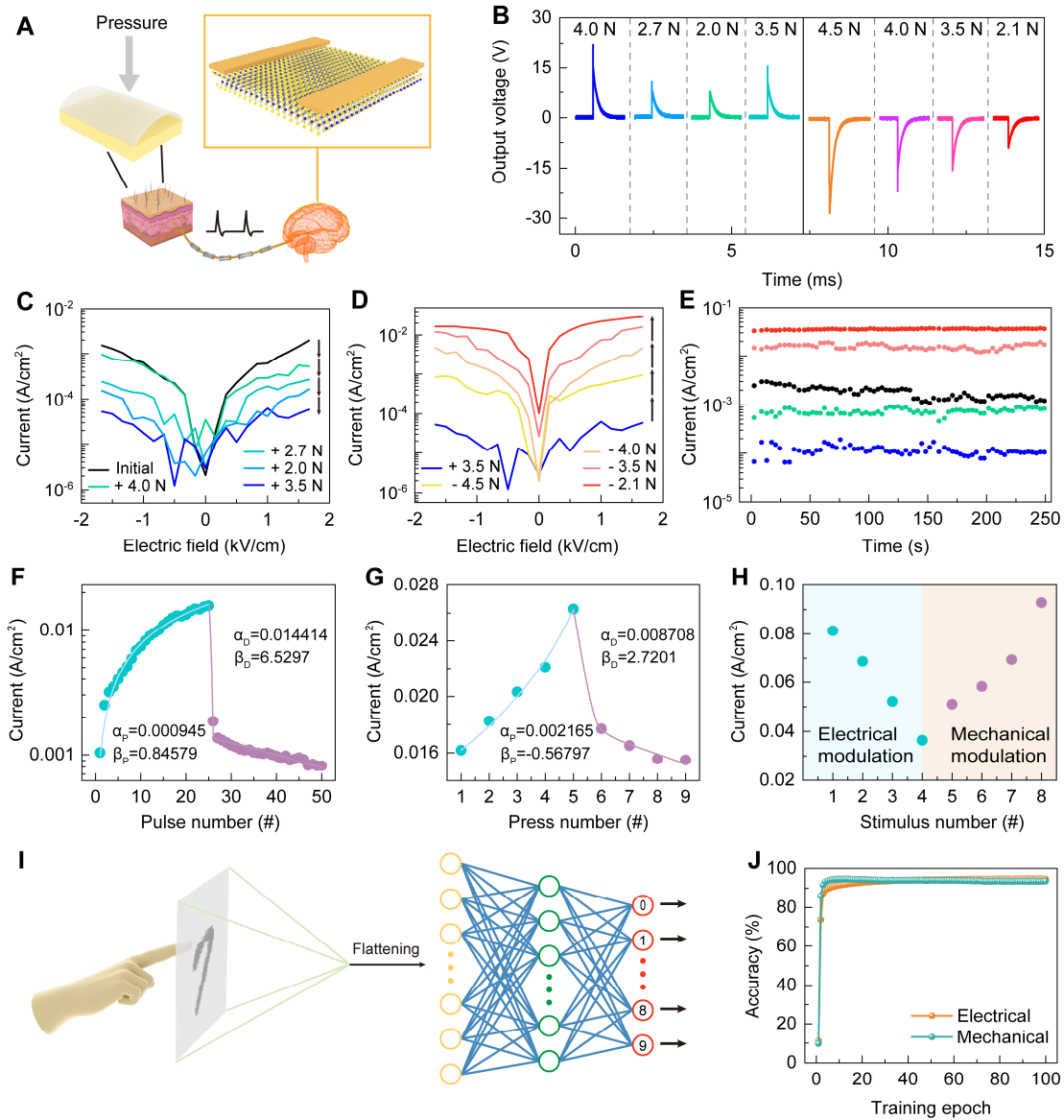

**Fig. 3. Mechanical switching of ferroelectric polarization for neuromorphic stress computing.**
**(A)** Schematic of neuromorphic stress system consisting of a pressing mode triboelectric unit and an α-In$_2$Se$_3$ ferroelectric memristor. The triboelectric unit, which directly receives press signals, can be used to simulate biological mechanoreceptor and the α-In$_2$Se$_3$ device can be used to mimic neural synapse. **(B)** Typical output voltages from the pressing mode triboelectric unit upon applying different press force. The output voltages are positive under forward connection, and vice versa. **(C)** Mechanical-force-dependent resistance switching for α-In$_2$Se$_3$ ferroelectric memristor under forward connection and **(D)** backward connection. **(E)** The retention property for different press-force-induced currents demonstrates the non-volatile memory effect under mechanical stresses. Read voltage: 1 V. **(F)** Potentiation and depression characteristics stimulated by consecutive ±2 V pulse voltages with a width of 10 ms. **(G)** Potentiation and depression characteristics stimulated by mechanical press force of approximately 0.5 N. To trigger the transition from potentiation to depression, the triboelectric unit was reversed from backward to forward connection. **(H)** Potentiation and depression realized by the electrical pulse voltage and the mechanical force modulation. Electrical pulse: 2 V with 10 ms; Mechanical force: 0.5 N under backward connection. **(I)** Illustration of artificial neural network used for handwritten digits image recognition. **(J)** Image recognition rates fed with LTP/LTD characteristics under electrical (green curve) and mechanical (orange curve) stimulations. The accuracy rates can reach 94.7 % and 93.6 % after 100 training cycles, respectively.

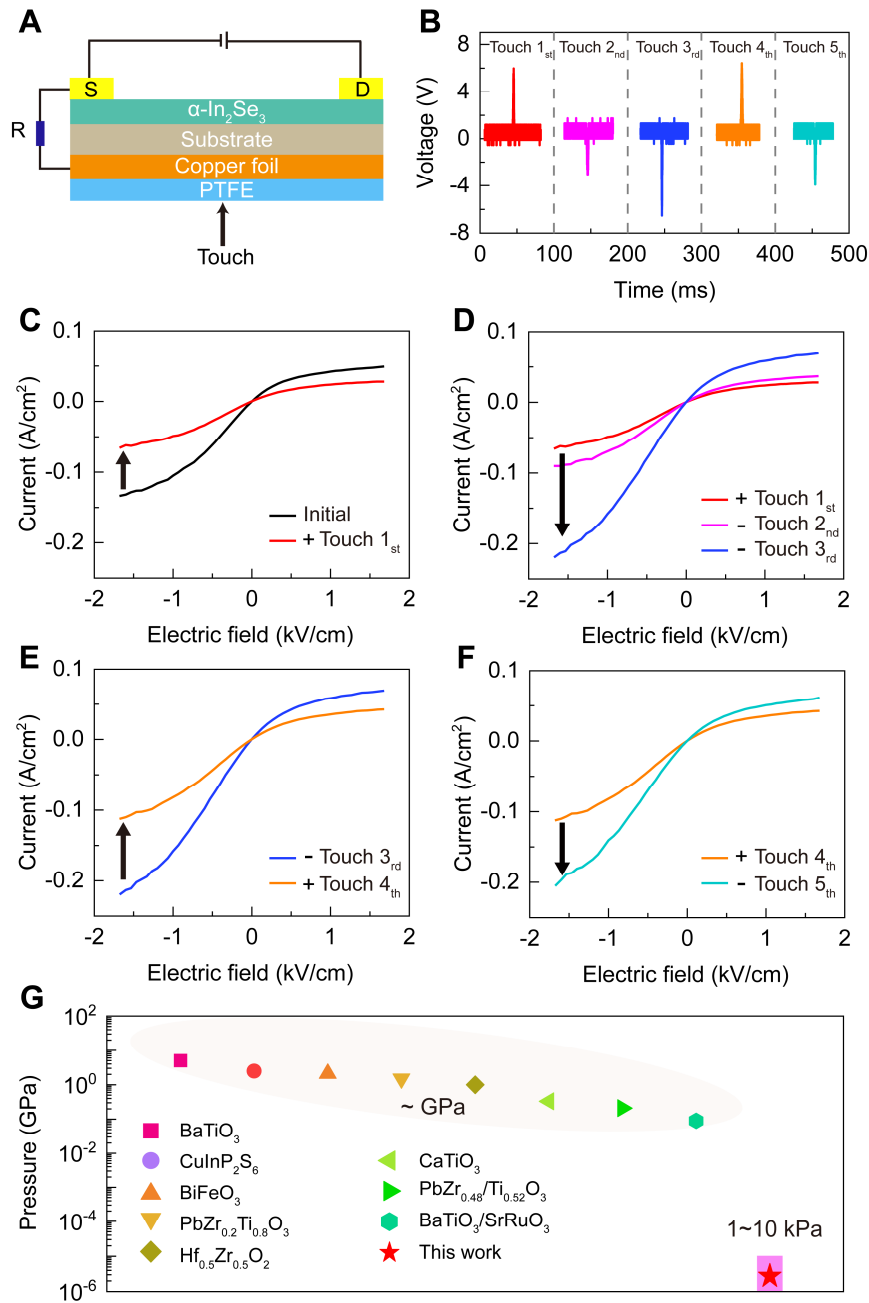

**Fig. 4. Tactile mechanical flipping of ferroelectric polarization using single-electrode mode triboelectric unit.** **(A)** Schematic of our devices for realizing tactile switching of ferroelectric polarization. Single-electrode triboelectric unit adhered with back side of Si substrate can be adopted to effectively sense tactile signals for the device. **(B)** Typical output voltages from triboelectric unit under forward and backward connections. Resistance switching arising from continuous touch of the single-electrode triboelectric unit: **(C)** Current decreases upon applying the first finger touch (+ Touch $1_{st}$) under the forward connection, while **(D)** it increases upon applying finger touch under backward connection. **(E)** and **(F)** reversible tactile-induced resistance switching. "+" and "−" represent forward and backward connections, respectively. **(G)** Comparison of mechanical pressures in different works for switching ferroelectric polarization. Our work exhibits a record-low mechanical pressure, ranging from 1~10 kPa.

# Supplementary Materials for

# Ultralow-pressure mechanical-motion switching of ferroelectric polarization

Baoyu Wang *et al*.

Corresponding author: Jian Sun, jian.sun@csu.edu.cn; Bin Yu, yu-bin@zju.edu.cn; Fei Xue, xuef@zju.edu.cn

**This PDF file includes:**

Supplementary Texts S1 to S6

Figs. S1 to S23

**Supplementary Text**

**Supplementary Text S1: Underlying physical mechanisms of triboelectricity-induced polarization switching**

The conversion of external mechanical force into electrical field through triboelectric units primarily relies on the contact electrification and charge transfer between material A and B. The polarization ($P_S$) of dielectric materials leads to the redistribution of free charges across the two electrodes and the potential difference between them builds a second electric field, which is the driving force for electron flow in the external circuit. The term $P_S$ comes from pre-existing electrostatic charges on the media and movement driven by external mechanical excitation. In turn, it can estimate the coupling effects of media polarization and energy conversion from mechanical excitation. The dielectric polarization caused by an external electric field is $P$ and the total electric field, denoted as $E$, is composed of the external electric field and the electric field generated by the dielectric materials polarization. Thus, the electric displacement vector ($D$) can be modified as:

$$D = \varepsilon_0 E + P + P_S = \varepsilon E + P_S = D' + P_S \tag{1}$$

where $\varepsilon_0$ and $\varepsilon$ is the permittivity vacuum and the relative dielectric constant, respectively. $D'$ is the traditional definition of the electric displacement vector. Using the newly defined $D$, the total displacement current ($J_D$) is expressed as:

$$J_D = \frac{\partial D'}{\partial t} + \frac{\partial P_S}{\partial t} = \varepsilon \frac{\partial E}{\partial t} + \frac{\partial P_S}{\partial t} \tag{2}$$

where $\varepsilon \frac{\partial E}{\partial t}$ represents the displacement current responsible for electromagnetic waves traveling at the speed of light in vacuum and $\frac{\partial P_S}{\partial t}$ represents the time-dependent dielectric materials polarization (*59*).

Considering that the charged dielectric medium can move at arbitrary velocity (inhomogeneous velocity *v (r, t)*), under external mechanical excitation the medium has a time-dependent volume, shape and boundary. Wang et al. proposed the expand

Maxwell's equations for the mechano-driven slow-moving medium system, which combine the mechano-electromagnetic interaction fields and reveal the complex dynamic electromagnetic phenomena of a moving medium (*60*). Using the newly defined **D** and inhomogeneous velocity **v** (**r**, *t*) to express the electrodynamics inside the media concerning the moving reference frame, the expand Maxwell's equations are:

$$\nabla \cdot \mathbf{D}' = \rho_f - \nabla \cdot \mathbf{P}_S \tag{3}$$

$$\nabla \cdot \mathbf{B} = \mathbf{0} \tag{4}$$

$$\nabla \times (\mathbf{E} + \mathbf{v}_r \times \mathbf{B}) = -\frac{\partial}{\partial t}\mathbf{B} \tag{5}$$

$$\nabla \times [\mathbf{H} - \mathbf{v}_r \times (\mathbf{D}' + \mathbf{P}_S)] = \mathbf{J}_f + \rho_f \mathbf{v} + \frac{\partial}{\partial t}(\mathbf{D}' + \mathbf{P}_S) \tag{6}$$

in which $\rho_f$ and $\mathbf{J}_f$ are the free charge density and the free current density, respectively. **B** is the magnetic field and **H** is the magnetizing field. The term $\mathbf{v}$ is the moving velocity of the reference frame and $\mathbf{v}_r$ is the relatively moving velocity with respect to the moving reference frame (*61*). The expanded Maxwell's equations integrate the electric field and electromagnetic phenomenon in a moving medium to explain how external mechanical excitation is converted into electrical energy from a physical perspective.

It is essential to establish an equivalent circuit model to illustrate how mechanical forces reverse ferroelectric polarization, as shown in fig. S1. Our used triboelectric units work at low frequencies (by manually sliding or pressing with the hand) under the quasi-static condition, so the generated time-dependent electric field changes slowly. The triboelectric units acting as the source can be considered as a series connection of a voltage source and a variable capacitor (*62*). The output voltage is:

$$V = -\frac{Q}{C(x)} + V_{OC}(x) \tag{7}$$

where *x* is the separation distance and *C(x)* represents the triboelectric units' capacitance. $Q$ and $V_{OC}$ are the output charge and open circuit voltage, respectively. The mechanical excitation is converted into an electric field through contact electrification and electrostatic induction. And then the induced charges are distributed at the terminals of the load resistor ($R_L$) and the ferroelectrics through the external circuit owing to the different potential between materials A and B. Finally, an electric field ($\mathbf{E}_{tribo}$) induced by the triboelectricity is applied on the ferroelectrics.

According to the Landau theory, in order to switch the ferroelectric polarization, an external electric field ($\mathbf{E}_{ext}$) should be applied to overcome the energy barrier ($E_F$), which is the energy difference between zero point and ± **P** states (*22*). It can be given by:

$$u(P) = u_{LG}^0 - \mathbf{P} \cdot \mathbf{E}_{eff} \tag{8}$$

$$E_{eff} = E_{ext} + E_{int} \tag{9}$$

in which $u(P)$ is the free-energy density and $u_{LG}^0$ represents Gibbs free-energy density at zero external electric field. $E_{int}$ refer to the local internal fields and $E_{eff}$ is the coupling field of the applied external electric field and the local internal field (*22*). The internal electric field includes depolarization field, ferroelastic strain and other associated fields. By integrating the triboelectric unit with the ferroelectric memristor, at this time, the $E_{tribo}$ is equivalent to $E_{ext}$ possessing sufficient energy to ensure that the $E_{eff}$ surpasses the energy barrier required for polarization switching.

## Supplementary Text S2: Discussion on the debate over α-In$_2$Se$_3$ in-plane polarization

We attempt to demonstrate triboelectricity-induced mechanical motion switching of ferroelectric polarization. This approach can sidestep a few hurdles from conventional flexoelectricity, such as huge pressure, and can be extended to other ferroelectric materials including BiFeO$_3$ (see fig. S23 for details).

For the material model used in this study (i.e., ferroelectric α-In$_2$Se$_3$), we have noticed the disputation of the presence of its in-plane polarization, as the in-plane crystal symmetry contradicts Neumann's principle. This debate is only supported by, to the best of our knowledge, one paper published in 2024(*63*). However, theoretical calculations and various experimental techniques—such as SHG, PFM and lateral electrical hysteresis measurements—have provided strong evidence supporting the existence of in-plane ferroelectric polarization in α-In$_2$Se$_3$ crystal, which constitutes a mainstream view on this debate.

Theoretically, to address the intuitive violation of Neumann's principle and explain the presence of α-In$_2$Se$_3$ in-plane dipole, recent studies have proposed a new concept of fractional quantum ferroelectricity(*64, 65*). Based on atomic structure analysis and theoretical calculations, the authors suggest that the in-plane polarization originates from the in-plane displacement of Se atoms between two energetically equivalent ferroelectric states, leading to a fractionally quantized polarization that breaks the *P3m1* symmetry of the system.

Experimentally, numerous prior works have used these characterization techniques to confirm the existence of α-In$_2$Se$_3$ in-plane polarization. For example, SHG measurement has demonstrated broken inversion symmetry in both out-of-plane and in-plane orientations, as revealed by the nonzero susceptibility tensor ratio between $\chi_{xxy}$ and $\chi_{zxx}$(*36*) (see the following interpretations for further details). Vector PFM has confirmed the persistence of both in-plane and out-of-plane polarization by simultaneously measuring the electromechanical responses from those perpendicular directions(*66*). Electrical hysteresis, depending on lateral or vertical device structure, has been considered strong evidence for in-plane or out-of-plane polarization switching(*67-69*). This is because the resistance hysteresis primarily originates from polarization charges-tuned interface barriers.

Inspired by these early works, in this paper we have conducted not only PFM measurements as in Fig. 2 and S4 but also systematic electrical hysteresis loop measurements as shown in Fig. 2, 3, and 4 to demonstrate the ferroelectric polarization switching. The underlying working mechanism of electrical hysteresis loops of our lateral ferroelectric memristor is also in-depth analyzed, particularly how ferroelectric polarization determines resistance states, as shown in Supplementary Text S6. Despite the possible crosstalk issues in vector PFM measurement, our electrical hysteresis collected from the lateral device structure further complements the PFM findings and provides additional evidence of polarization switching.

Therefore, we believe that we have provided solid evidence via the combination of PFM and electrical measurements to demonstrate mechanical motion switching of polarization. Our study represents a big leap in pushing this effect to device applications.

**Interpretations of SHG in α-In$_2$Se$_3$**: building on our previous work, we next explain why SHG results can indicate both in-plane and out-of-plane non-centrosymmetry. The hexagonal α-In$_2$Se$_3$ possesses eleven second-order susceptibilities for the whole χ tensor(*70*), but three out of eleven are fully independent, i.e., $\chi_{xxy}$, $\chi_{zxx}$ and $\chi_{zzz}$. Theoretical nonlinear responses from the in-plane and out-of-plane orientations are plotted in fig. S5A. Supposed that only in plane susceptibilities ($\chi_{xxy}$) take effect, the acquired SHG response will show a C6 symmetry. By contrast, if only out-of-plane contributions ($\chi_{zxx}$ and $\chi_{zzz}$) exist, the produced SHG signal will be like a circle. However, as both in-plane and out-of-plane contributions occur, the resulting SHG signal will display a C3 symmetry owing to the interference. Experimentally, SHG responses from α-In$_2$Se$_3$ crystal display remarkable C3 symmetry (fig. S5B), thus implying the coexistence of in-plane and out-of-plane asymmetry.

**Supplementary Text S3: Operating principle of triboelectric units**

In our experiment, three distinct types of triboelectric units were utilized: free-standing (sliding) mode, pressing mode, and single-electrode mode. Their operating principles are shown in fig. S6 and interpreted below.

Sliding mode: As shown in fig. S6A, once the cooper foil from top movable square contacts with PTFE film, the PTFE film with electronegativity captures negative triboelectric charges and the cooper foil is positively charged, resulting in negative and positive charges are distributed in the left and right cooper foils of bottom component, respectively (*39*). During the sliding process, charge transfer occurs owing to the potential difference between two copper foils, thereby generating a pulse voltage in the external circuit.

Pressing mode: As shown in fig. S6B, positive and negative charges are induced by ITO and PDMS respectively under separation (*54*). When pressing upper PET film, the top stacking (ITO/PDMS) and bottom stacking (ITO) have contact electrification, and thus generate a pulse voltage. Note that each press can generate a pulse voltage.

Single-electrode mode: For the single-electrode mode (fig. S6C), contact electrification and charge transfer occur between our positive charged finger and the PTFE film upon touching (*41*). In the case of separation, positive charges are electrostatically induced and distributed on the copper foil, resulting in a pulse voltage in the external circuit.

The waveform of the output voltage is mainly determined by the motion velocity (*71*). Under the same pressure, the contact area between the two triboelectric layers with positive and negative charges is equivalent, leading to the same amount of transferred triboelectric charges. According to Equation 10, the charge $\boldsymbol{Q}$ is the integral of current $\boldsymbol{I}$ (i.e., $\boldsymbol{U/R}$) over time $\boldsymbol{t}$. When the motion velocity increases (i.e., $\boldsymbol{t}$ decreases), the peak current $\boldsymbol{I}$ becomes larger with a smaller pulse width accordingly. Since the output voltage is equal to the current multiplied by the load resistance, the voltage increases as well. Therefore, the pulse width shortens as the velocity is raised.

$$\boldsymbol{Q} = \int \frac{\boldsymbol{U}}{\boldsymbol{R}} \, d\boldsymbol{t} \tag{10}$$

When the velocity is constant, the pulse width also remains unchanged. Increasing the pressure will lead to a higher amplitude of the pulse voltage.

**Supplementary Text S4: Measurement setup of neuromorphic stress device**

Our measurement setup is shown in fig. S10. A pressure gauge measures the mechanical stress pressure, an oscilloscope records the corresponding output voltages, and a source meter measures the conductance variation of the memristor. The output voltages under various mechanical press forces are shown in fig. S11A, and the amplitude of output voltage as a function of the applied force is shown in fig. S11B.

**Supplementary Text S5: Measurement of LTP/LTD characteristics under mechanical press stimulation**

The characterization of LTP/LTD under mechanical press stimulation is as follows. First, the mechanical press is applied to modulate the resistance state. Subsequently, as depicted in fig. S15A, a sweeping electric field ranging from -1.6 to 1.6 kV/cm is applied to measure the corresponding current states. Finally, the current values at 1.6 kV/cm are extracted for analysis, as shown in the enlarged section of fig. S15B.

**Supplementary Text S6: Mechanism of α-In$_2$Se$_3$ memristor resistance variation**

The Schottky barrier height at the interface of the electrode and channel of the α-In$_2$Se$_3$ memristor is crucial to the enhancement of on/off ratio, which is also indicative of the underlying working mechanism. During our experiments, we have fabricated three types of devices, i.e., dual Ohmic contact electrodes, and single electrode Schottky contact. We used triboelectric unit to modulate their resistance states with the same experimental condition, in order to further explain the working mechanism of our devices.

We deposited Cr/Au electrodes on the α-In$_2$Se$_3$ layer via thermal evaporation to form Ohmic contacts, and measured its electrical properties as a field effect transistor (FET) to confirm the Ohmic contacts. As shown in fig. S19A, high-conductivity silicon is used as the back gate and the 285 nm SiO$_2$ is the dielectric layer. fig. S19B depicts the transfer curves at $V_{DS}$ = 1 V but with sweeping $V_{GS}$ from -60 to 60 V. It is worth noting that the off-state current of the FET reaches the $10^{-9}$ A (nA) level, which may be attributed to the incompletely depleted electrons in the channel with a thick oxide layer (285 nm). The output curves shown in fig. S19C verify the good Ohmic contacts at the source and drain electrodes. Then, we measured the device's performance as a ferroelectric memristor. fig. S19D shows the electrical hysteresis loop under bidirectional sweeping voltages. It can be seen that although the ferroelectric polarization is completely reversed by applying a large voltage, only a small window appears on the left side, which is due to the lack of an adjustable Schottky barrier. Correspondingly, as shown in fig. S19E, the variation of the current is minor under the modulation of triboelectric unit: only a slight change on the left side. The energy band diagram shown in fig. S19F illustrates that without the Schottky barrier, the transport of electrons will not be affected, and the current will not change even though the ferroelectric polarization is reversed.

In addition, we also found another two devices with Schottky contact but at only one terminal. The Schottky contact may originate from impurities or adsorbed molecules introduced at the interface between the electrode and the α-In$_2$Se$_3$ layer during the microfabrication process. fig. S20A and D show that these two devices have Schottky contacts on the right and left sides respectively, and Ohmic contacts on the counter sides. Under the excitation of the triboelectric unit, only the currents of both Schottky contact sides are notably changed while the currents of the other sides of Ohmic contact are barely changed, as shown in fig. S20B and E. The underlying mechanism is shown in the energy band diagram in fig. S20C and F: The reversal of ferroelectric polarization in α-In$_2$Se$_3$ forms an opposite ferroelectric polarization electric field, which causes an increase or decrease in the energy band and results in a notable variation in the Schottky barrier and then the current. On the contrary, even if the energy band on the ohmic contact side is bent under the ferroelectric switching, there is a lack of adjustable Schottky barrier, meaning that the movement of electrons is not affected either at the $P_{up}$ or $P_{down}$ state, so the current will not be affected.

In order to better explain the relationship between the Schottky barrier and currents, the variation of the Schottky barrier heights for the three aforementioned devices (i.e., ohmic contact, left Schottky contact and right Schottky contact) are demonstrated through calculations. It can be calculated using the thermionic emission model (*31*):

$$J = A^*T^2 exp\left(-\frac{e\emptyset_s}{kT}\right) exp\left(\frac{eV}{kT} - 1\right) \tag{11}$$

Note that the ferroelectric polarization is induced by external mechanical press rather than by applying sweep voltage (*V*) on α-In$_2$Se$_3$ devices, so the corresponding current density *J* induced by *V* can be ignored. Then the equation is expressed as:

$$J = A^*T^2 exp\left(-\frac{e\emptyset_s}{kT}\right) \tag{12}$$

where $A^*$ is the effective Richardson constant, $k$ is the Boltzmann constant and $\emptyset_s$ represents the Schottky barrier height. Considering we focus on the variation of the Schottky barrier height ($\Delta\emptyset_s$) and the other parameters are constant, it can be extracted from a simpler formula as follows (*72*):

$$ln\,[I_{HRS}/I_{LRS}] \sim -\Delta\emptyset_s/KT \tag{13}$$

in which $I_{HRS}$ and $I_{LRS}$ represent the current under high resistance state (P$_{up}$) and low resistance state (P$_{down}$) modulated by mechanical press. The dependence relationship between the variation of Schottky barrier height ($\Delta\emptyset_s$) and the modulation of current is shown in fig. S21, indicating that larger variations in the Schottky barrier height correspond to greater modulations in the current.

Furthermore, we investigated the dependence of the current state variations on the pulse width of the stimulus voltages. The large hysteresis in fig. S22A resulting from ferroelectric polarization switching indicates the Schottky contacts. Pulse voltages of ± 10V with pulse widths of 10 μs and 1 μs were applied to the device. As shown in fig. S22B, a notable change in current is observed with the 10 μs pulse width, while fig. S22C shows no apparent change with the 1 μs pulse width. These results suggest that in our device structure, the pulse width below 10 μs is insufficient to deliver the energy required for ferroelectric polarization switching.

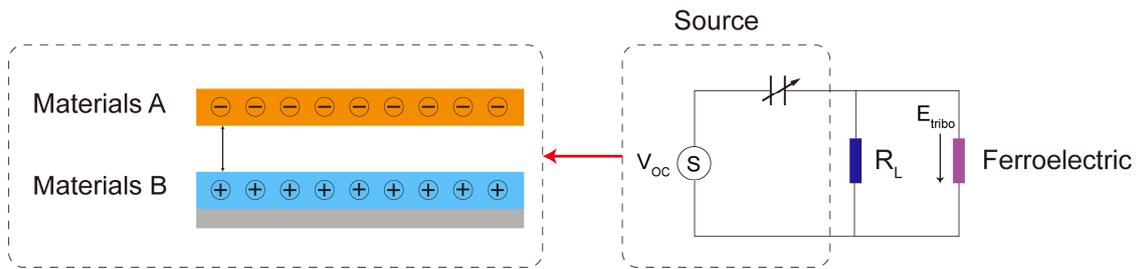

**Fig. S1. Equivalent circuit model for triboelectric units and ferroelectric materials.** The triboelectric units can be considered as a power source, where external mechanical excitation is converted into an electric field that is applied to the ferroelectric material, leading to polarization switching.

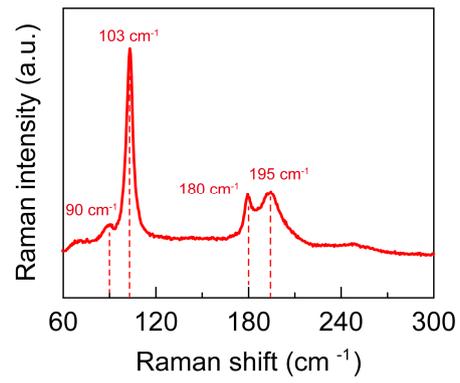

**Fig. S2. Raman spectra of the used α-In$_2$Se$_3$ sample.** The located peak at 90 cm$^{-1}$ indicates hexagonal 2H phase.

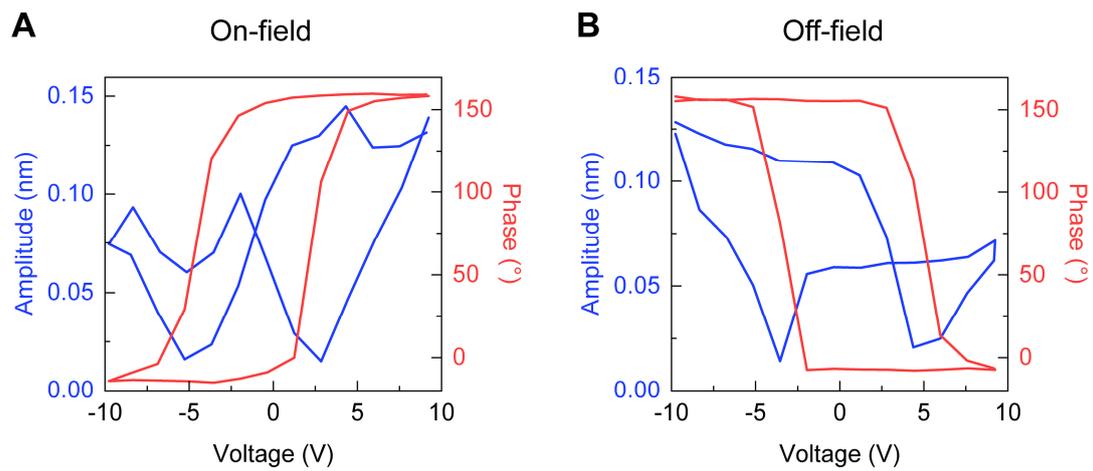

**Fig. S3. Local PFM phase and amplitude hysteresis loops. (A), (B)** The on-field (A) and off-field (B) hysteresis loops.

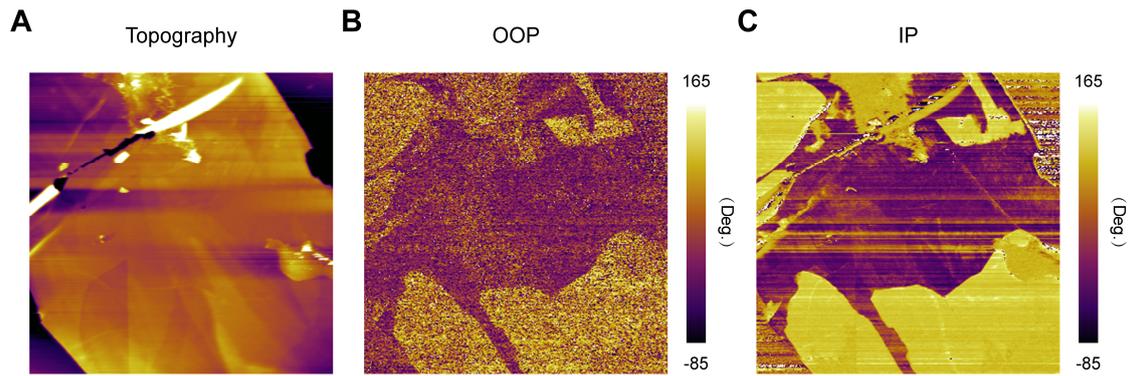

**Fig. S4. Spontaneous OOP and IP domains of our used α-In$_2$Se$_3$ sample. (A)** The topography of measured α-In$_2$Se$_3$ layer. **(B)** Spontaneous OOP polarization. **(C)** Spontaneous IP polarization. It can be observed that both OOP and IP polarizations are independent of the morphology.

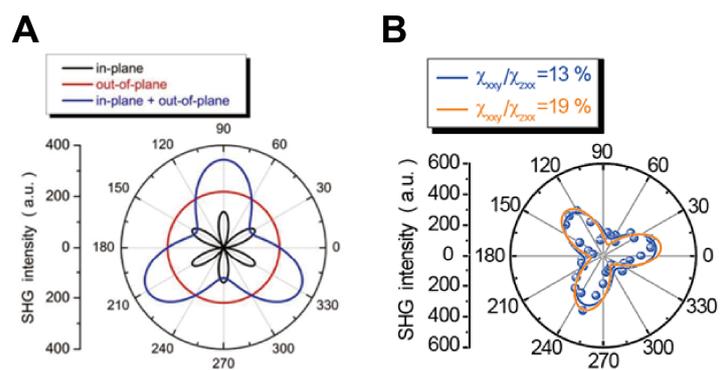

**Fig. S5. Theoretical and experimental SHG signal of hexagonal α-In$_2$Se$_3$.** **(A)** Theoretical responses from pure in-plane or out-of-plane directions. **(B)** Experimental SHG collected from a α-In$_2$Se$_3$ flake. These data were reproduced from our paper published in *ACS Nano*[36].

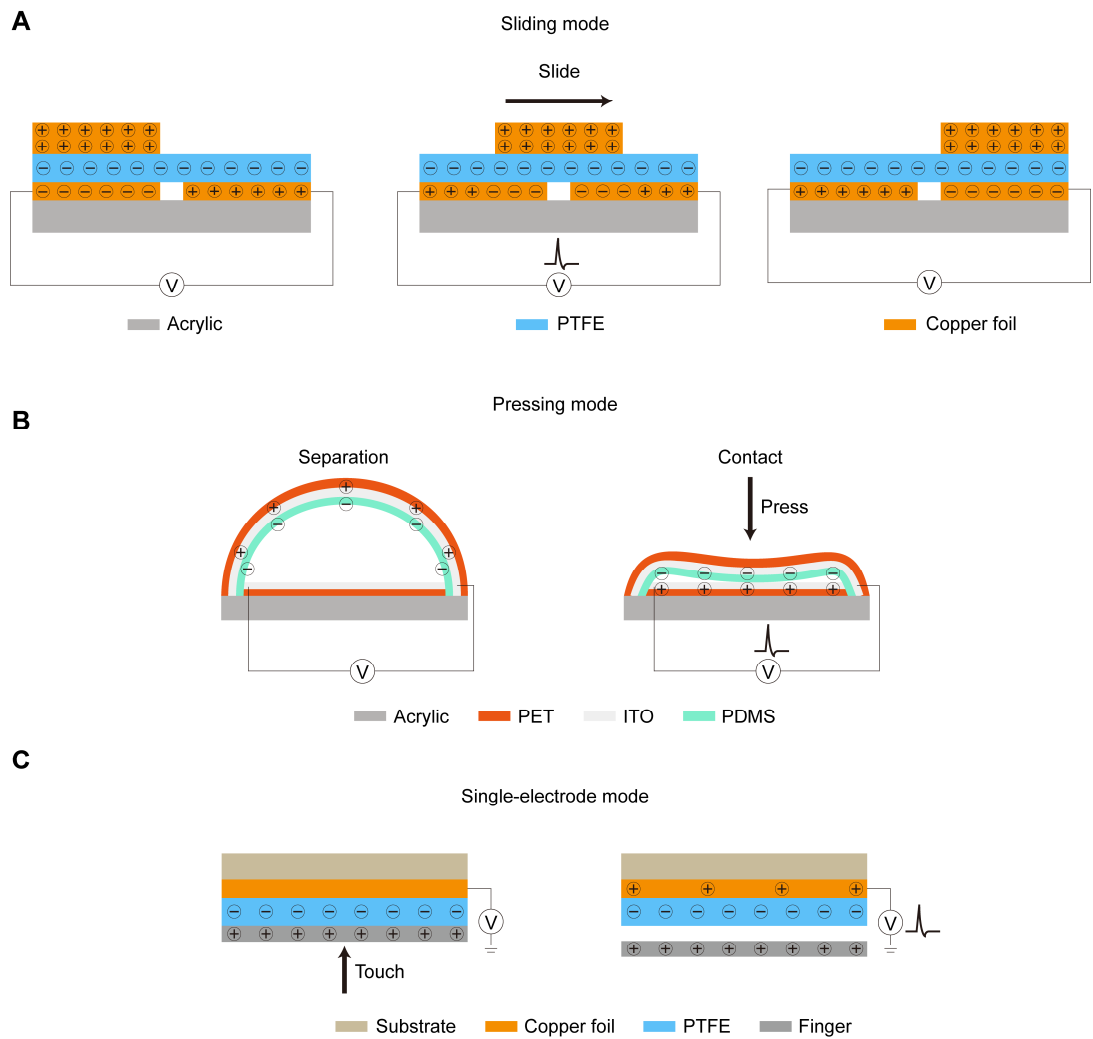

**Fig. S6. The operating principle of three distinct types of triboelectric units used in our experiments. (A)** The pulse voltage is generated once the upper copper foil slides on the PTFE film. **(B)** By pressing the upper triboelectric layer, the PDMS and the bottom ITO come into contact, resulting in the pulse voltage in the external circuit. **(C)** Our positive charged finger can directly touch the PTFE film to generate the pulse voltage.

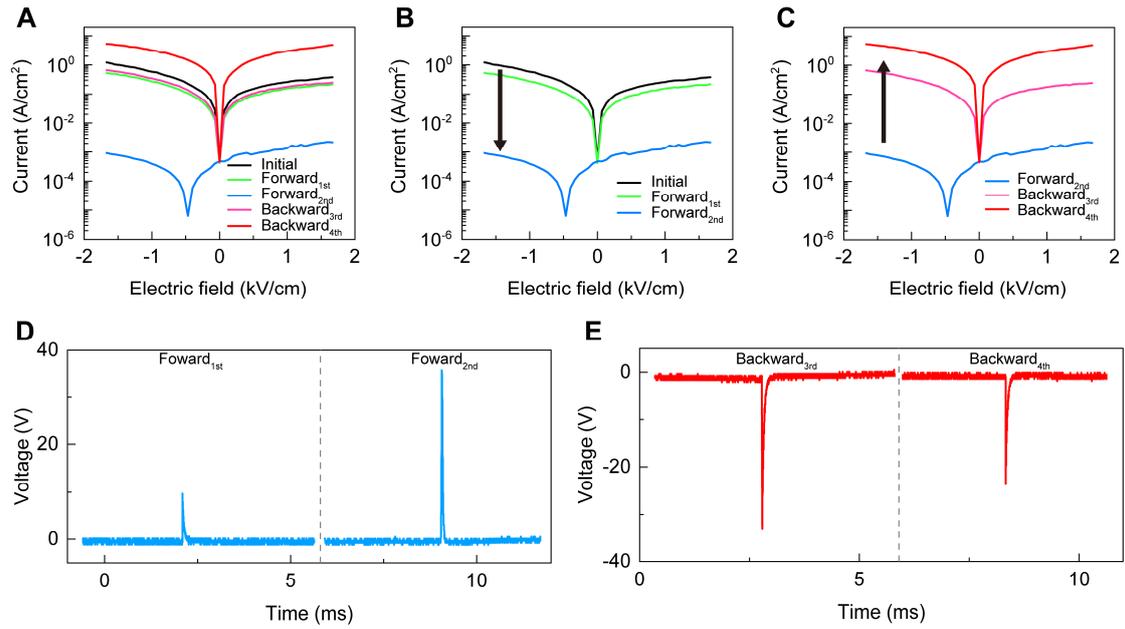

**Fig. S7. Modulating α-In$_2$Se$_3$ memristor current state by a sliding-mode triboelectric unit. (A)** The initial and modulated resistance switching under small read sweeping electric fields from -1.6 ~ 1.6 kV/cm. **(B)** Resistance variation with two times mechanical friction under forward connection. Black curve represents the initial current state. Green and blue curves represent the resistance state after the first (Forward$_{1st}$) and the second (Forward$_{2nd}$) sliding under forward connection, respectively. **(C)** Pink and red curves represent resistance states after the third and the fourth sliding under backward connection, respectively. The output pulse voltages corresponding to sliding triboelectric units under **(D)** forward and **(E)** backward connections, respectively. The amplitude is dependent on the applied friction force.

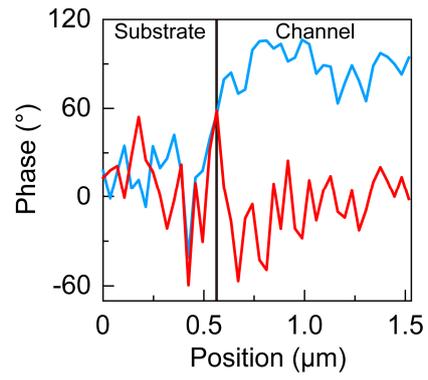

Fig. S8. Comparison of piezo-response phases taken from the blue and red lines in forward and backward phase mapping, respectively.

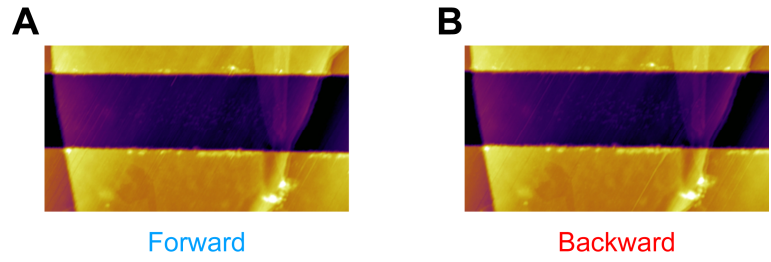

Forward     Backward

**Fig. S9. Corresponding topography after poling. (A), (B)** The topography after forward (A) and backward (B) poling. The unchanged topography rules out the possibility of ferroelectric polarization switching being caused by morphological changes.

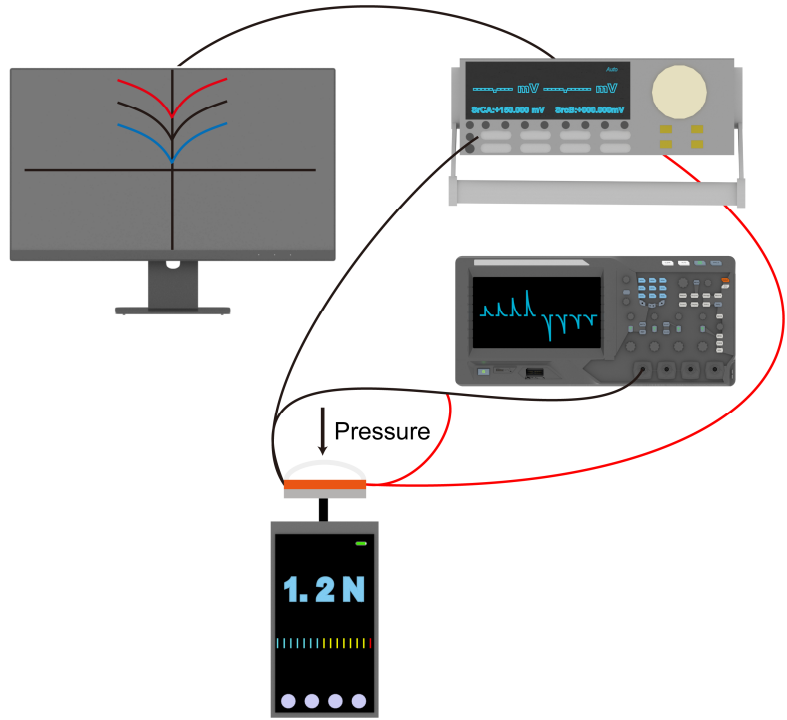

**Fig. S10. Schematic diagram of our measurement setup.** A pressing mode triboelectric unit is placed on a pressure gauge to record the stress magnitude. An oscilloscope and 2636 source meter are connected in parallel with the memristor to measure the mechanical output voltages and resistance switching, respectively.

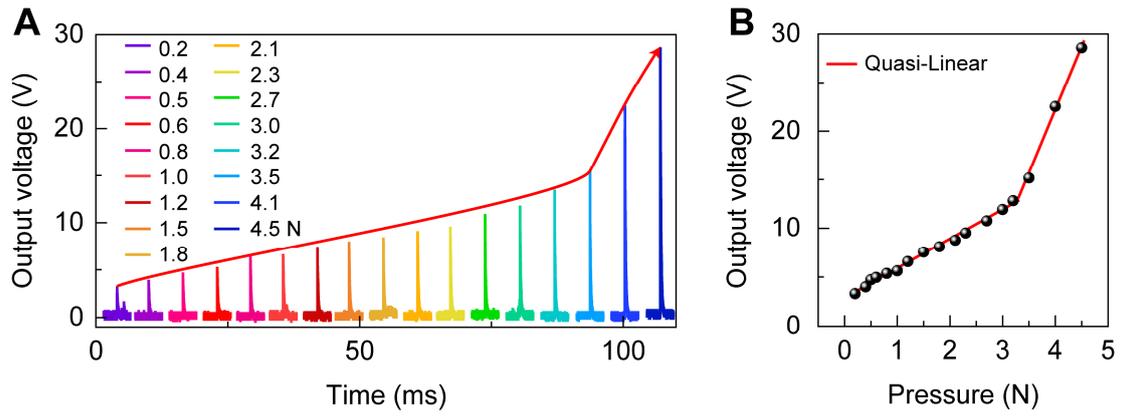

**Fig. S11. Output voltages of pressing mode triboelectric unit under various applied mechanical pressures. (A)** Output voltages increase with increasing pressures. **(B)** The dependence of output voltage amplitude on pressure exhibits a quasi-linear relationship.

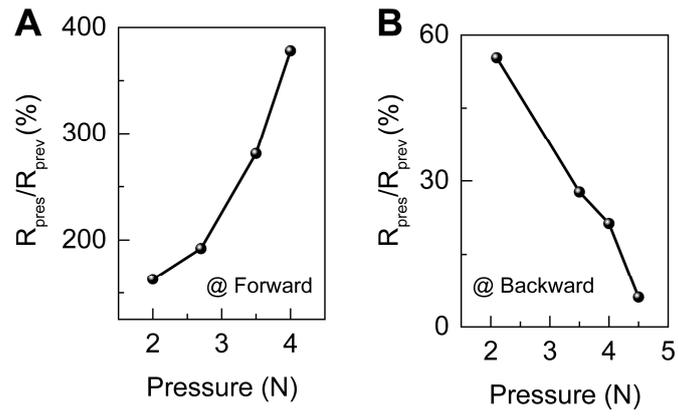

**Fig. S12. Pressure-dependent resistance variation.** **(A)** and **(B)** The resistance variation ($R_{pres}/R_{prev}$) modulated by mechanical pressures under forward and backward connections, respectively.

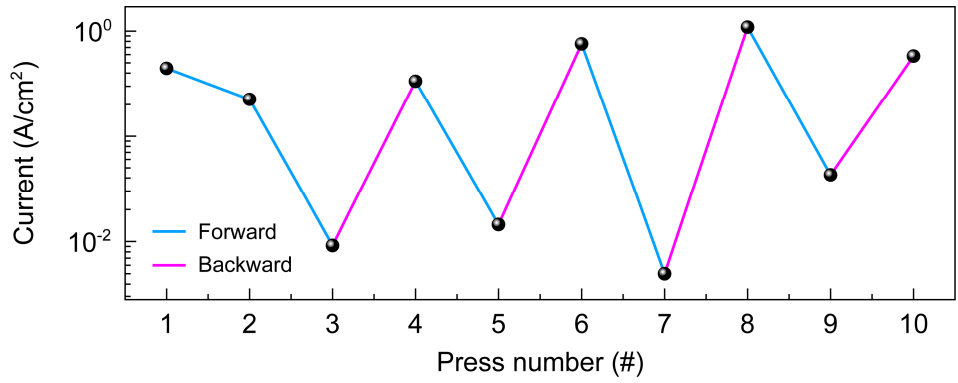

**Fig. S13. Another device's current state is repeatedly modulated by triboelectric unit through changing the electrical connections.** The Blue and pink curves correspond to the forward and backward connections with triboelectric units, respectively. Note that the nonuniform variation is owing to the different pressure applied each time.

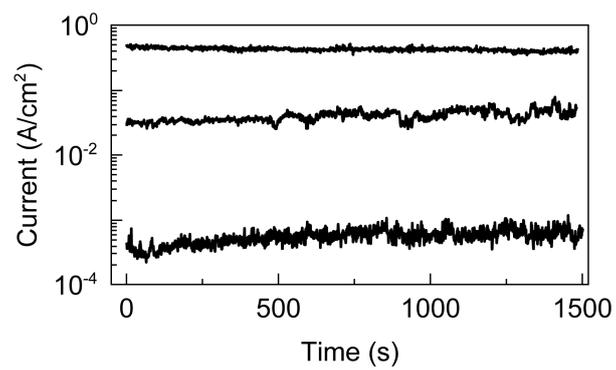

**Fig. S14. Three current states maintain for 1500 s.**

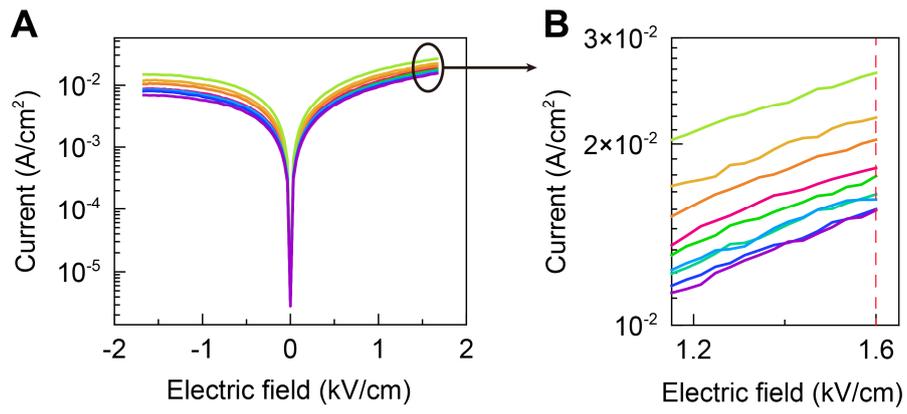

**Fig. S15. Measurement process of LTP/LTD under mechanical press stimulation. (A)** Read current states after each application of mechanical press. **(B)** Extract the current values at 1.6 kV/cm electric field.

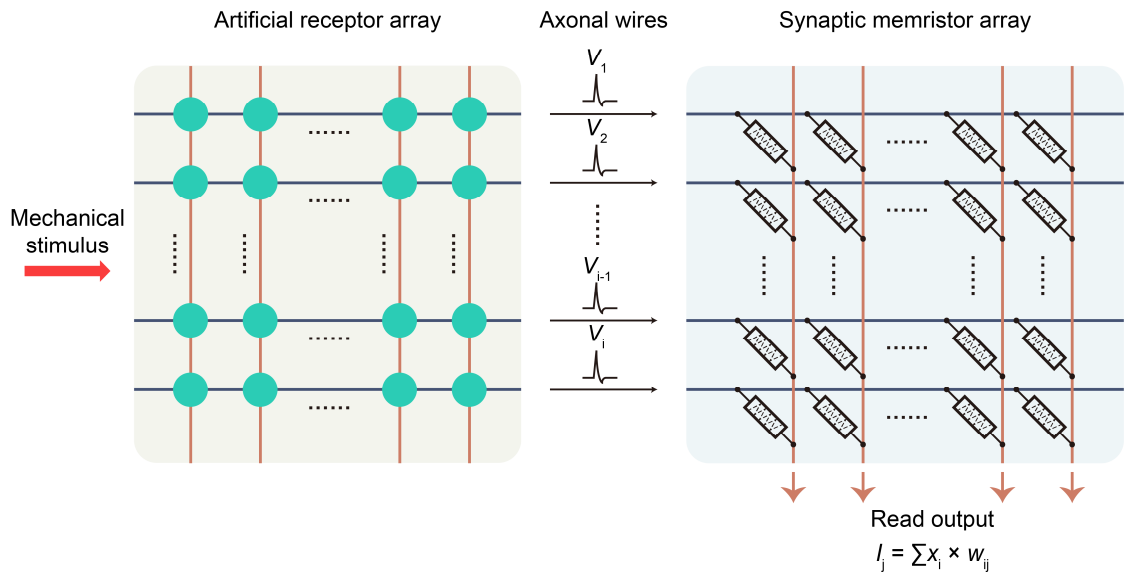

**Fig. S16. Schematic of a conceptual neuromorphic tactile system based on the large array of α-In$_2$Se$_3$ devices.** This system incorporates triboelectric units that transform mechanical stimuli into analog receptor potentials, forming an artificial receptor array. The axonal wires connect the receptor array and memristor array, acting as nerve afferents to transmit the action potential, analogous to how information is conveyed in biological systems. The α-In$_2$Se$_3$ devices array functions as synaptic elements, enabling the sense, storage, and processing of tactile information.

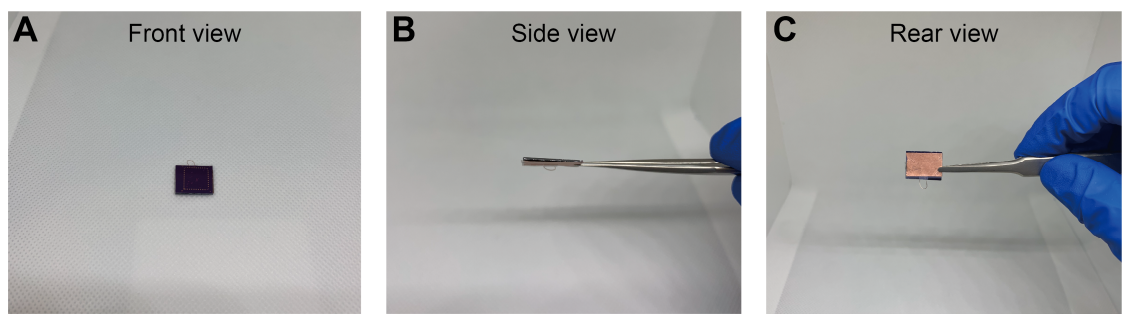

**Fig. S17. The real photo of single-electrode mode triboelectric unit integrated with α-In$_2$Se$_3$ devices. (A)** Front view. **(B)** Side view. **(C)** Rear view.

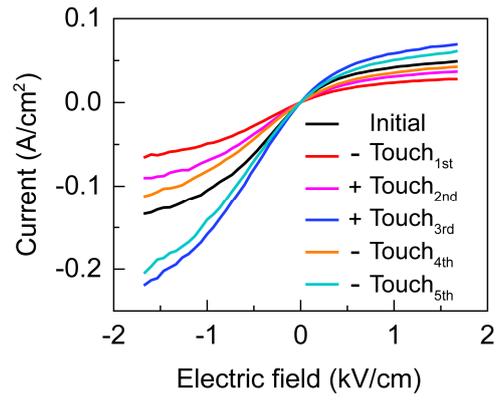

**Fig. S18. Modulating the α-In$_2$Se$_3$ memristor resistance state by touching the single-electrode mode triboelectric unit.**

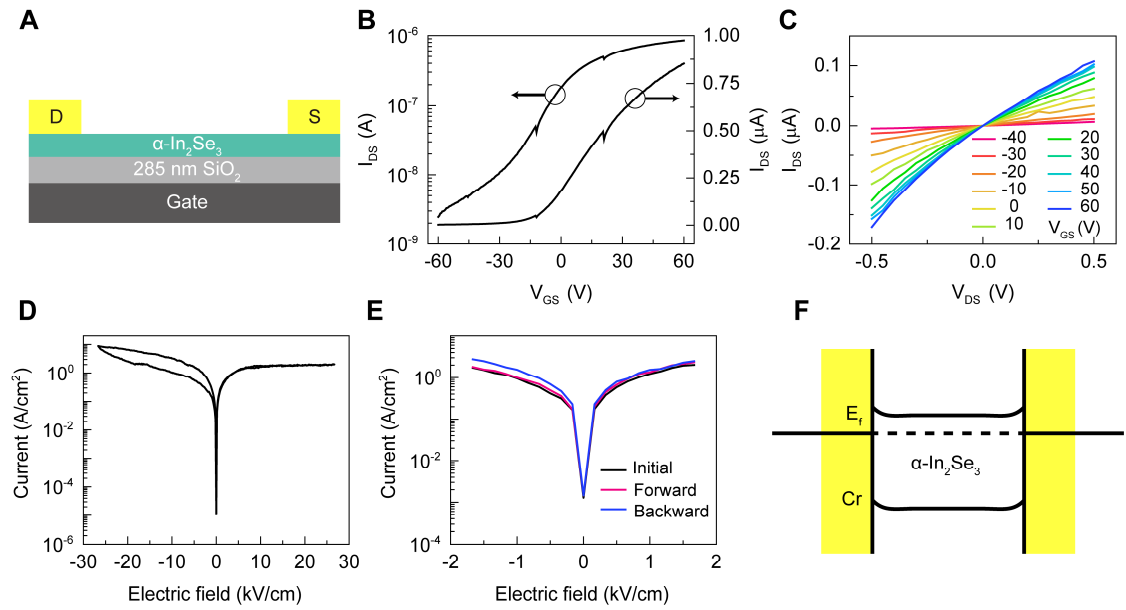

**Fig. S19. Ohmic-contact α-In₂Se₃ mem-transistor. (A)** Schematic of α-In₂Se₃ ferroelectric field transistor, in which high doping Si is the back gate and α-In₂Se₃ acts as the channel with Ohmic contacts. **(B)** Transfer curves of an α-In₂Se₃ transistor at $V_{DS}$ = 1 V. The left and right are the logarithmic and linear coordinates, respectively. **(C)** Output curves of the α-In₂Se₃ transistor at $V_{GS}$ from -40V to 60V. The linear I-V curves at different $V_{GS}$ indicate good Ohmic contact across the α-In₂Se₃ channel. **(D)** I-V electrical hysteresis loops under 26 kV/cm electric field sweeping. The sweeping direction is from zero to the positive maximum voltage and then back to the negative maximum voltage. Note that the loop is small due to the low Schottky barrier. **(E)** α-In₂Se₃ memristor current modulated by the triboelectric unit. Black curve is the initial state. Red and blue curves correspond to the forward and backward connections, respectively. It can be seen that the current is almost unregulated. **(F)** Energy band diagram of the α-In₂Se₃ device. The Schottky barrier is low.

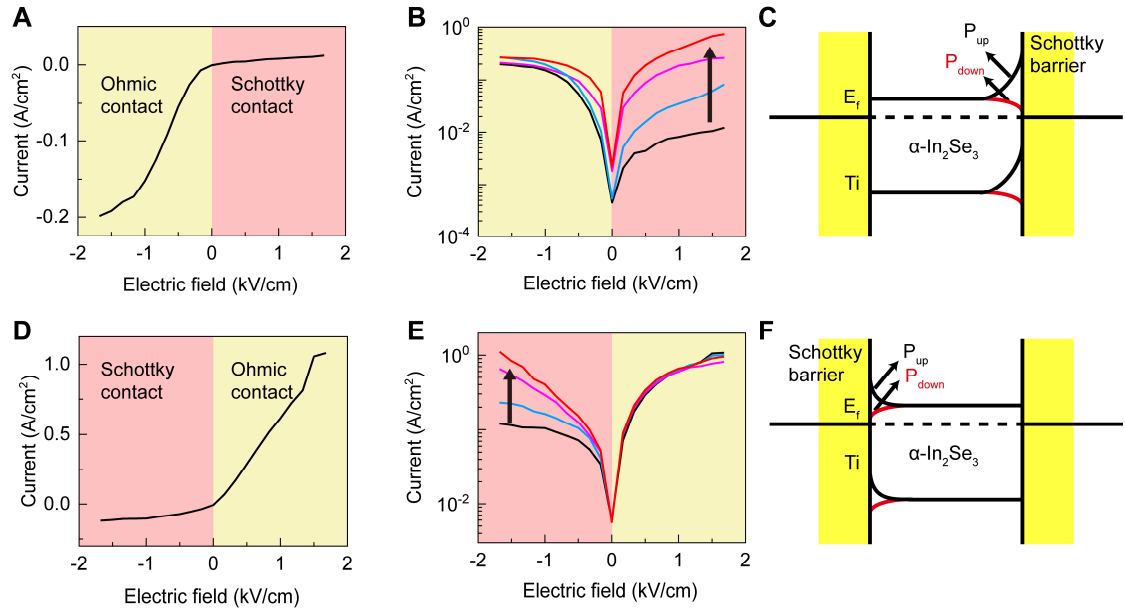

**Fig. S20. Schottky contact at only one side. (A)** Single Schottky contact at the right side, and the counter side is an Ohmic contact. **(B)** The current state at the Schottky contact side (right) is obviously modulated by the triboelectric unit, and the other side is almost unchanged. **(C)** The height of the Schottky barrier at the right side is changed notably under ferroelectric polarization switching, making the energy band bend upward or downward, corresponding to the $P_{up}$ and $P_{down}$ states, respectively. **(D)** Single Schottky contact at the left side, and the counter side is an Ohmic contact. **(E)** The current state at the Schottky contact side (left) is obviously modulated by TENG, and the other side is almost unchanged. **(F)** The height of the Schottky barrier at the left side is changed notably under ferroelectric polarization switching, making the energy band bend upward or downward, corresponding to the $P_{up}$ and $P_{down}$ states, respectively.

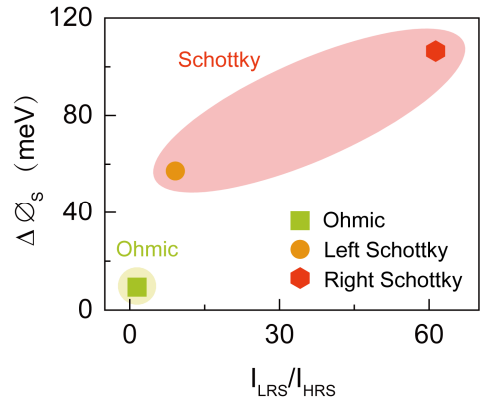

**Fig. S21. The dependence relationship between the variation of Schottky barrier height ($\Delta \phi_s$) and current change.** The result indicates that a larger variation in the Schottky barrier height corresponds to greater modulations in the current.

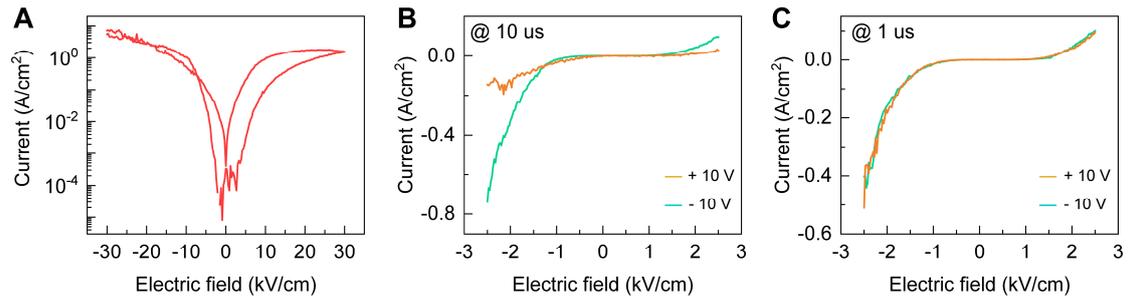

**Fig. S22. Current state variations under ± 10V pulse voltage stimulation with different pulse widths. (A)** Electrical hysteresis curve of the device originating from ferroelectric polarization switching. **(B)** Current states after stimulation of 10 μs pulse voltage. **(C)** Current states after stimulation of 1 μs pulse voltage.

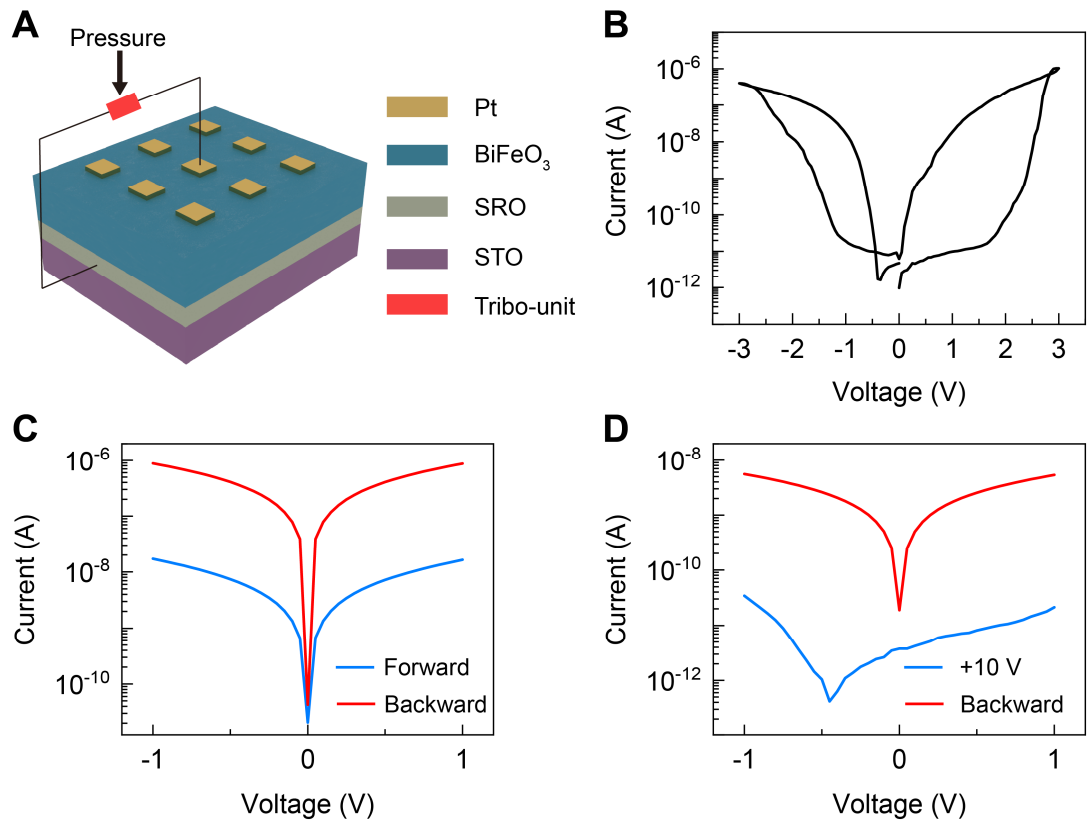

**Fig. S23. Triboelectricity-induced mechanical switching of BiFeO₃ ferroelectric polarization.** **(A)** Structure of BiFeO₃ memristor, in which Pt and SRO are top and bottom electrodes, respectively. **(B)** I-V sweeping loops of the BiFeO₃ memristor. **(C)** Mechanically modulated resistance switching by the triboelectric effect. Blue and red curves indicate the current states after sliding the triboelectric unit under forward and backward connections, respectively. **(D)** Resistance switching with the modulation of electrical voltage and mechanical force. Blue and red curves are the current states after applying +10 V voltage and sliding the triboelectric unit under backward connection, respectively.